\documentclass[a4paper,12pt]{article}
\usepackage{amsmath}
\usepackage{amsfonts,amssymb}
\usepackage{preprint}
\begin{document}
\numberwithin{equation}{section}
\setlength{\voffset}{-1cm}
\setlength{\baselineskip}{15pt}
\rm
%
\def\calF{{\cal F}}
\def\gothg{\mathfrak{g}}
\def\tB{\skew5\widetilde{B}{}}
\def\bC{\bar{C}{}}
\def\lag{{\cal L}}
\def\wightman#1{\langle\,#1\,\rangle}
\def\twightman#1{\wightman{#1}_{\hbox{\scriptsize T}}}
\def\bdel{\hbox{\boldmath$\delta$}}
\def\hj{\skew5\widehat{j}{}}
\def\jB{j_{\hbox{\tiny B}}}
\def\hjB{\hj_{\hbox{\tiny B}}}
\def\hQ{\widehat{Q}{}}
\def\QB{Q_{\hbox{\tiny B}}}
\def\hQB{\hQ_{\hbox{\tiny B}}}
%
\hrule height0pt depth0pt
\vspace*{-12pt}
\rightline{\bf RIMS-1341}
\vspace*{10pt}
\vspace*{50pt}
\centerline{
\LARGE\bfseries 
Exact Solutions to the Two-dimensional BF and}
\vskip15pt
\centerline{
\LARGE\bfseries 
Yang-Mills Theories in the Light-cone Gauge
}
\vskip70pt
\renewcommand{\thefootnote}{\fnsymbol{footnote})}
\centerline{\large
 Mitsuo Abe\footnote{E-mail: \abemail}
}
\centerline{\it Research Institute for Mathematical Sciences, 
 Kyoto University,}
\centerline{\it Kyoto 606-8502, Japan}
\vskip10pt
\centerline{\large
 Noboru Nakanishi\footnote{Professor Emeritus of Kyoto University. 
 E-mail: \nnmail}
}
\centerline{\it 12-20 Asahigaoka-cho, Hirakata 573-0026, Japan}
\renewcommand{\thefootnote}{\alph{footnote})}
\setcounter{footnote}{0}
\vskip15pt
\vskip80pt
\centerline{\itshape\bfseries Abstract}
\vskip5pt
It is shown that the BRS-formulated two-dimensional BF theory in the 
light-cone gauge (coupled with chiral Dirac fields) is solved very 
easily in the Heisenberg picture.  The structure of the exact solution 
is very similar to that of the BRS-formulated two-dimensional quantum
gravity in the conformal gauge. In particular, the BRS Noether charge 
has anomaly. Based on this fact, a criticism is made on the reasoning
of Kato and Ogawa, who derived the critical dimension $D=26$ of string
theory on the basis of the anomaly of the BRS Noether charge. By adding
the $\tB^2$ term to the BF-theory Lagrangian density, the exact 
solution to the two-dimensional Yang-Mills theory is also obtained. 
\vfill\eject
%
%
%
\section{Introduction}
During the last decade, we have developed a new method of solving 
quantum field theory in the Heisenberg 
picture\cite{AN-QEG,AN-QED,AN-L,AN-BF,AN-CP} 
and explicitly constructed exact solutions to various two-dimensional 
models.\cite{AN-2DQG,AN-Z,AN-W1,AN-W2,AN-W3,AN-LG2DQG,AN-CG2DQG} 
Our method is different from the conventional covariant perturbation 
theory or the Feynman path-integral approach in the following respects:
\begin{itemize}
\item[1.] 
In our approach, we first find the (nonlinear) infinite-dimensional
Lie algebra of field operators and then construct its representation 
in terms of Wightman functions. On the other hand, path-integral 
approach directly gives the solution in terms of Green's functions 
without making any consideration at the operator level.
\item[2.] 
The Green's function, which is a vacuum expectation value of
T$^*$-product ($=$ Lagrangian-formalism version of T-product) of 
operators, does {\it not\/} respect field equations and, therefore, 
Noether theorem, in contrast to the Wightman function ($=$ vacuum 
expectation value of simple product of operators).
\item[3.] 
When there is a fundamental field having no canonical conjugate, 
the notion of T$^*$-product becomes {\it indefinable\/} at the 
operator level. Then the qualitative argument based on Green's 
functions may become quite misleading, especially concerning 
conservation laws and anomaly. On the other hand, our approach gives 
the solution consistent with Noether theorem even in such a case.
\item[4.] 
In our approach, some {\it nonlinear\/} field equations may happen to 
be slightly violated at the representation level because 
singular products involved in them may not be faithfully represented.
This phenomenon is called 
``field-equation anomaly''.\cite{AN-W3} The same one does not exist in 
path-integral approach, but, instead, some unusual loop diagrams arise 
from the difference between T$^*$-product and T-product, and they cause
anomaly.\cite{AN-PO}
\item[5.] 
If field-equation anomaly is present for a field equation $\calF=0$, 
the existence or non-existence of anomaly for a particular symmetry 
$G$ {\it can\/} depend on the expressions for the current of $G$ 
which are different from each other by an operator multiple of $\calF$.
In path-integral approach, this phenomenon cannot be seen; what one 
can do is to convert the anomaly encountered in $G$ into the anomaly 
of another symmetry.\cite{Takahashi}
\end{itemize}
\par
In our previous paper published already eight years ago,\cite{AN-BF} 
we constructed the exact solution to the BRS-formulated two-dimensional
BF theory in the Landau gauge.  The analysis is rather complicated and 
does not encourage further investigation. In the present paper, we 
consider the same model in a non-covariant gauge.  
Since in the axial-gauge case, we encounter a trouble in quantizing FP
ghosts, we discuss the BRS-formulated two-dimensional BF theory 
{\it in the light-cone gauge\/} (we abbreviate it as LGBF) coupled 
with $D$ chiral Dirac fields. The analysis can easily be extended to 
the case of the two-dimensional Yang-Mills theory. 
\par
We find that the exact solution to LGBF can be obtained very simply 
and that its  structure is quite similar to the BRS-formulated 
two-dimensional quantum gravity {\it in the conformal 
gauge\/}\cite{AN-CG2DQG} (we abbreviate it as CGQG). 
In particular, in both models, all field equations except for the 
equation involving B-field are linear, and the B-field equation 
exhibits a field-equation anomaly (for $D\not=26$ in CGQG and for 
$D\not=0$ in LGBF).  Although these models are formulated to be 
BRS-invariant (that is, an anomaly-free BRS generator exists), the 
BRS {\it Noether\/} charge exhibits BRS anomaly because of the 
mechanism explained above (Item 5).
\par
In 1983, Kato and Ogawa,\cite{KO} who found BRS anomaly for $D\not=26$ 
in CGQG (with boundary conditions corresponding to an open string), 
claimed that the critical dimension $D=26$ of a string could be 
obtained in the BRS-formulated string theory.  Recently, we have 
criticized their claim by pointing out that the reason why they 
encountered BRS anomaly is merely due to their adoption of the BRS 
{\it Noether\/} charge as the BRS generator, that is, the model itself 
is strictly BRS-invariant for any value of $D$.\cite{AN-KO}  
The result found in the present paper strengthens our criticism: If 
the Kato-Ogawa reasoning of deriving the critical dimension were 
applied to LGBF, one would obtain $D=0$, a result which is meaningless.
Thus the adequacy of deriving the critical dimension of string theory 
on the basis of BRS anomaly is quite questionable.
\par
In Sec.\ 2, we present the operator algebra of LGBF.
In Sec.\ 3, we construct a complete set of Wightman functions and find 
the existence of field-equation anomaly for the B-field equation.
In Sec.\ 4, the BRS invariance of the model is confirmed, but it is 
shown that the BRS Noether charge exhibits anomaly for $D\not=0$.
In Sec.\ 5, we solve the BRS-formulated two-dimensional Yang-Mills 
theory. The final section is devoted to discussion.
\vskip30pt
%
%
%
\section{Operator algebra}
We discuss the BRS-formulated two-dimensional BF theory in the 
light-cone gauge. The fields considered are a non-abelian gauge field 
$A_\mu{}^a$, where Lie algebra $\gothg$ is characterized by structure 
constants $f^{abc}$, the conjugate field $\tB^a$, the B-field $B^a$, 
the FP-ghost $C^a$, the FP-antighost $\bC^a$, and $D$ chiral Dirac 
fields $\psi_M$. 
With light-cone coordinates $x^\pm=(x^0 \pm x^1)/\sqrt{2}$, 
the Lagrangian density is given 
by
\begin{eqnarray}
\lag &=& 
 \tB^a(\partial_-A_+{}^a-\partial_+A_-{}^a-f^{abc}A_+{}^bA_-{}^c)
 + B^a A_-{}^a  \nonumber\\
 &&
 + i\bC^a(\delta^{ab}\partial_- + f^{acb}A_-{}^c)C^b 
 + i\psi_M{}^\dagger(\partial_- - iA_-{}^aT^a)\psi_M,      \label{lag}
\end{eqnarray}
where $T^a$ denotes the representation matrix of $\gothg$, normalized 
as tr$(T^aT^b)=\frac{1}{2}\delta^{ab}$.
\par
Field equations derived from \eqref{lag} are
\begin{eqnarray}
&&A_-{}^a=0,                                               \label{fe1}\\
&&\partial_-\varPhi=0\ \ \hbox{for}\ \ 
  \varPhi=A_+{}^a,\, \tB^a,\, C^a,\,\bC^a,\,\psi_M,        \label{fe2}\\
&&B^a + \partial_+ \tB^a + f^{abc}(A_+{}^b\tB^c-i\bC^bC^c)
  +\psi_M{}^\dagger T^a \psi_M = 0.                        \label{fe3}
\end{eqnarray}
From \eqref{fe3} and \eqref{fe2}, we obtain $\partial_- B^a=0$.
Thus all fields are functions of $x^+$ only.
\par
Canonical quantization is carried out by taking $A_+{}^a$, $C^a$ and 
$\psi_M$ as canonical variables.  We then obtain
\begin{eqnarray}
&& [\tB^a(x), \, A_+{}^b(y)]=-i\delta^{ab}\delta(x^+-y^+),\label{os1}\\
&& \{ \bC^a(x), \, C^b(y)\}=\delta^{ab}\delta(x^+-y^+),   \label{os2}\\
&& \{ \psi_M(x), \, \psi_N{}^\dagger(y)\}=\delta_{MN}\delta(x^+-y^+).
                                                          \label{os3}
\end{eqnarray}
Because of the $x^-$-independence, \eqref{os1}--\eqref{os3} are 
already two-dimensional (anti)commuta-tion relations.
All other (anti)commutators vanish if the B-field $B^a$ is not 
involved. Thus without $B^a$, the model is a free-field theory.
\par
In the operator-level calculation, we can regard \eqref{fe3}, which 
is a unique nonlinear equation, as the defining equation of $B^a$.  
By using it, we calculate commutation relations involving $B^a$:
\begin{eqnarray}
&& [B^a(x), \, A_+{}^b(y)]
    =i(\delta^{ab}\partial_+ + f^{acb}A_+{}^c(x))\delta(x^+-y^+), \label{os4}\\
&& [B^a(x), \, \varPhi^b(y)]
    =-if^{abc}\varPhi^c(x)\delta(x^+-y^+) \ \ \hbox{for} \ \ 
      \varPhi^a=\tB^a, \, C^a,\,\bC^a,\,B^a,                      \label{os5}\\
&& [B^a(x), \, \psi_M(y)]=T^a\psi_M(x)\delta(x^+-y^+).            \label{os6}
\end{eqnarray}
From \eqref{os4}--\eqref{os6}, we see that the local-gauge commutation 
relations, proposed by Kanno and Nakanishi in 1985,\cite{KN} are 
satisfied without non-gauge terms.
\par
The $\delta'$-term of \eqref{os4} can be eliminated by introducing a field
\begin{equation}
 B'{}^a \equiv B^a + \partial_+\tB^a.                        \label{B'}
\end{equation}
We then find 
\begin{eqnarray}
&& [\varPhi^a(x), \, B'{}^b(y)]=-if^{abc}\varPhi^c(y)\delta(x^+-y^+)
  \ \ \hbox{for} \ \ \varPhi^a=A_+{}^a,\,\tB^a,\,C^a,\,\bC^a,\,B'{}^a,
                                                     \qquad \label{os7}\\
&& [\psi_M(x),\, B'{}^b(y)]=-T^b\psi_M(y)\delta(x^+-y^+).   \label{os8}
\end{eqnarray}
\par
It is now easy to calculate all multiple commutators. 
For example, we have
\begin{equation}
[\,[ A_+{}^a(x),\,B'{}^b(y)],\,\tB^c(z)]
 = f^{abc}\delta(x^+-y^+)\delta(y^+-z^+).
\end{equation}
\vskip20pt
%
%
%
\section{Wightman functions}
As stated in the Introduction, we construct the representation of the 
operator algebra found in Sec.\ 2 in terms of Wightman functions, 
so as to be consistent with all vacuum expectation values of multiple 
commutators and to satisfy the energy-positivity condition.\cite{AN-W1}
\par
All 1-point functions are arbitrary in principle, but it is natural to 
set them equal to zero.
\par
Nonvanishing 2-point and 3-point functions (apart from those obtained 
by field permutations) are as follows:
\begin{eqnarray}
&&\wightman{A_+{}^a(x_1)\tB^b(x_2)}
  =\frac{1}{2\pi}\delta^{ab}\frac{1}{x_1{}^+ - x_2{}^+ - i 0},    
                                                       \label{W-AtB}\\
&&\wightman{C^a(x_1)\bC^b(x_2)}
  = - \frac{i}{2\pi}\delta^{ab}\frac{1}{x_1{}^+ - x_2{}^+ - i 0}, 
                                                       \label{W-CbC}\\
&&\wightman{\psi_M(x_1)\psi_N{}^\dagger(x_2)}
  = - \frac{i}{2\pi}\delta_{MN}\frac{1}{x_1{}^+ - x_2{}^+ - i 0}; 
                                                       \label{W-psipsi} 
\end{eqnarray}
\samepage{
and
\begin{eqnarray}
&&\wightman{A_+^a(x_1)B'{}^b(x_2)\tB^c(x_3)}
  = -f^{abc}\varphi_3(x_1{}^+,x_2{}^+,x_3{}^+),        \label{W-ABtB}\\
&&\wightman{C^a(x_1)B'{}^b(x_2)\bC^c(x_3)}
  = if^{abc}\varphi_3(x_1{}^+,x_2{}^+,x_3{}^+),        \label{W-CBbC}\\
&&\wightman{\psi_M(x_1)B'{}^b(x_2)\psi_N{}^\dagger(x_3)}
  = \delta_{MN}T^b\varphi_3(x_1{}^+,x_2{}^+,x_3{}^+)   \label{W-psiBpsi} 
\end{eqnarray}
with
\begin{eqnarray}
&&\varphi_3(x_1{}^+,x_2{}^+,x_3{}^+)
  \equiv \frac{1}{(2\pi)^2}\cdot
         \frac{1}{(x_1{}^+-x_2{}^+-i0)(x_2{}^+-x_3{}^+-i0)}.
\end{eqnarray}
Generally, the nonvanishing truncated $n$-point functions\footnote{As 
for 2-point and 3-point functions, truncated and nontruncated are the 
same because all 1-point functions vanish.} are
\begin{eqnarray}
&&\twightman{A_+{}^a(x_1)B'{}^{b_2}(x_2)\cdots 
               B'{}^{b_{n-1}}(x_{n-1})\tB^c(x_n)}   \nonumber\\
&&\quad
 = i\twightman{C^a(x_1)B'{}^{b_2}(x_2)\cdots 
               B'{}^{b_{n-1}}(x_{n-1})\bC^c(x_n)}   \nonumber\\
&&\quad
 =(-1)^n\sum_{P(j_2,\ldots,j_{n-1})}^{(n-2)!}
   f(ab_{j_2}d_{j_2})f(d_{j_2}b_{j_3}d_{j_3})\cdots
   f(d_{j_{n-2}}b_{j_{n-1}}c)                       \nonumber\\
&&\hspace*{200pt}
    \times\varphi(x_1{}^+,x_{j_2}{}^+,\ldots,x_{j_{n-1}}{}^+,x_n{}^+)
\end{eqnarray}
with $f(abc)\equiv f^{abc}$ and
\begin{eqnarray}
&&\twightman{\psi_M(x_1)B'{}^{b_2}(x_2)\cdots 
            B'{}^{b_{n-1}}(x_{n-1})\psi_N{}^\dagger(x_n)}   \nonumber\\
&&\quad
 =-i^{n-1}\delta_{MN}\sum_{P(j_2,\ldots,j_{n-1})}^{(n-2)!}
   T(b_{j_2})T(b_{j_3})\cdots T(b_{j_{n-1}})
    \varphi(x_1{}^+,x_{j_2}{}^+,\ldots,x_{j_{n-1}}{}^+,x_n{}^+)\qquad
\end{eqnarray}
with $T(b)\equiv T^b$, where $P(j_2,\ldots,j_{n-1})$ denotes 
a permutation of $(j_2,\ldots,j_{n-1})$ and
\begin{eqnarray}
&&\varphi(x_1{}^+,x_{j_2}{}^+,\ldots,x_{j_{n-1}}{}^+,x_n{}^+) \nonumber\\
&&\qquad
 \equiv \frac{1}{(2\pi)^{n-1}}
        \frac{1}{(x_1{}^+-x_{j_2}{}^+ - i0)
                 (x_{j_2}{}^+-x_{j_3}{}^+ \mp i0) \cdots
                 (x_{j_{n-1}}{}^+-x_n{}^+ - i0)} \qquad   
\end{eqnarray}
with
\begin{eqnarray}
x_j{}^+ - x_k{}^+ \mp i0 
 &=& x_j{}^+ - x_k{}^+ - i0 \ \ \hbox{if} \ \ j<k \nonumber\\
 &=& x_j{}^+ - x_k{}^+ + i0 \ \ \hbox{if} \ \ j>k.        \label{xxmp0}
\end{eqnarray}
\par
We rewrite \eqref{fe3} into $\calF^a\equiv B'{}^a - F^a = 0$ with
\begin{equation}
F^a \equiv -f^{abc}(A_+{}^b\tB^c-i\bC^bC^c)
           -\psi_M{}^\dagger T^a \psi_M.                  \label{feaF}
\end{equation}%
\nopagebreak{}%
Then, using the generalized normal product rule, we obtain
\begin{eqnarray}
\wightman{B'{}^a(x_1) B'{}^b(x_2)}&=&0,                   \label{fea1}\\
\wightman{B'{}^a(x_1) F^b(x_2)}
  &=&\wightman{F^a(x_1) F^b(x_2)}                         \nonumber\\
  &=&-\frac{D}{2(2\pi)^2}\cdot
      \frac{\delta^{ab}}{(x_1{}^+ - x_2{}^+ - i0)^2}.     \label{fea2}\\
&&\nonumber
\end{eqnarray}
}
\vfill\break
\samepage{
\noindent
Thus we encounter field-equation anomaly for $D\not=0$.
\par
The perturbation-theoretical counterpart of this fact is as follows.
In spite of \eqref{fe1}, the Feynman propagator 
$\wightman{\hbox{T}^* B^a(x_1) A_-{}^b(x_2)}$ does {\it not\/} vanish 
but proportional to $\delta^2(x_1-x_2)$.
Therefore, the $B'$ self-energy diagrams, which are loop diagrams of 
$A_+\tB$, of $\bC C$ and of $\psi_M\psi_M{}^\dagger$, are nonvanishing.
The former two cancel, but the last one remains. This implies violation
of BRS invariance in perturbation theory.
\vskip20pt
%
%
%
\section{BRS invariance}
\nopagebreak{}%
The BRS transformation is given by
\begin{eqnarray}
&&\bdel A_\pm{}^a = \partial_\pm C^a + f^{acb}A_\pm{}^c C^b, \\
&&\bdel \tB^a = -f^{abc}C^b\tB^c, \\
&&\bdel C^a = -\frac{1}{2}f^{abc}C^bC^c, \\
&&\bdel \bC^a = iB^a = i(B'{}^a-\partial_+\tB^a),\\
&&\bdel B^a = 0, \quad \bdel B'{}^a = -f^{abc}\partial_+(C^b\tB^c),\\
&&\bdel \psi_M = iC^a T^a\psi_M, \quad
  \bdel \psi_M{}^\dagger = -i\psi_M{}^\dagger C^a T^a.
\end{eqnarray}
Of course, field equations and (anti)commutation relations are 
consistent with BRS invariance.
\nopagebreak{}%
\par%
In contrast to perturbation theory, our exact solution is consistent 
with BRS invariance.  Indeed, explicit calculation based on Wightman 
functions shows
\begin{eqnarray}
&&
\wightman{\bdel(A_+{}^a(x_1)\bC^b(x_2))}
 = \partial_+{}^{x_1}\wightman{C^a(x_1)\bC^b(x_2)}
   -i\partial_+{}^{x_2}\wightman{A_+{}^a(x_1)\tB^b(x_2)}=0,    \\
&&
\wightman{\bdel(A_+{}^a(x_1)\bC^b(x_2)\tB^c(x_3))}    \nonumber\\
&&\qquad
=  f^{aed}\wightman{A_+{}^e(x_1)\tB^c(x_3)}
          \wightman{C^d(x_1)\bC^b(x_2)} 
    + i\wightman{A_+{}^a(x_1)B'{}^b(x_2)\tB^c(x_3)}   \nonumber\\
&&\qquad\qquad 
   + f^{cde}\wightman{A_+{}^a(x_1)\tB^e(x_3)}
            \wightman{\bC^b(x_2)C^d(x_3)}             \nonumber\\ 
&&\qquad= 0, \\
&&
\wightman{\bdel(C^a(x_1)\bC^b(x_2)\bC^c(x_3))}        \nonumber\\
&&\qquad
= \frac{1}{2}f^{ade}\bigg[
       \wightman{C^d(x_1)\bC^b(x_2)}\wightman{C^e(x_1)\bC^c(x_3)}
        - (x_2\,\leftrightarrow\,x_3,\ b\,\leftrightarrow\,c)\bigg]
     \nonumber\\
&& \qquad\qquad 
  - i \wightman{C^a(x_1)B'{}^b(x_2)\bC^c(x_3)}
  + i \wightman{C^a(x_1)\bC^b(x_2)B'{}^c(x_3)}        \nonumber \\
&&\qquad=0, \\
&&
\wightman{\bdel(\psi_M(x_1)\bC^b(x_2)\psi_N{}^\dagger(x_3))} \nonumber\\
&&\qquad
 = i\wightman{C^a(x_1)\bC^b(x_2)}T^a
    \wightman{\psi_M(x_1)\psi_N{}^\dagger(x_3)}      
  +i\wightman{\psi_M(x_1)B'{}^b(x_2)\psi_N{}^\dagger(x_3)}   \nonumber\\
&&\qquad\qquad 
  +i\wightman{\bC^b(x_2)C^a(x_3)}
    \wightman{\psi_M(x_1)\psi_N{}^\dagger(x_3)}T^a           \nonumber\\
&&\qquad
 =0.
\end{eqnarray}
\par
The BRS Noether current $\jB{}^\mu$ is given by
\begin{eqnarray}
\jB{}^+ & = & 0, \\
\jB{}^- & = & \tB^a\partial_+C^a + f^{acb}\tB^a A_+{}^c C^b
               + \frac{1}{2}if^{abc}\bC^aC^bC^c 
               - C^a\psi_M{}^\dagger T^a \psi_M.
\end{eqnarray}
}
\eject\noindent%
We define another BRS current $\hjB{}^\mu$ by 
\begin{eqnarray}
\hjB{}^+ &\equiv& \jB{}^+,  \\
\hjB{}^- &\equiv& \jB{}^- + C^a\calF^a \nonumber\\
         &=& B^a C^a 
           - \frac{1}{2}if^{abc}\bC^aC^bC^c + \partial_+(\tB^aC^a),
\end{eqnarray}
which is, of course, equal to $\jB{}^\mu$ at the operator level because 
$\calF^a=0$.  But $\jB$ and $\hjB$ are not identical at the 
representation level because, as shown by \eqref{fea1} and \eqref{fea2}, 
$\calF^a\equiv B'{}^a-F^a=0$ exhibits field-equation anomaly.
\par
The BRS generators $\QB$ and $\hQB$ are defined by
\begin{eqnarray}
&& \QB = \int dx^+ \jB{}^-, \\
&& \hQB = \int dx^+ \hjB{}^-.
\end{eqnarray}
By using (anti)commutation relations but without using field
equations, we can confirm that
\begin{equation}
i[\,\hQB, \, \varPhi\,]_\mp=\bdel(\varPhi)               \label{hQ-Phi}
\end{equation}
for any fundamental field $\varPhi$. Likewise, we have
\begin{equation}
[\calF^a(x), \, \varPhi(y)]=0 \ \ \hbox{for}\ \ 
    \varPhi=A_+{}^b,\,\tB^b,\,C^b,\,\bC^b,\,\psi,        \label{F-Phi}
\end{equation}
but
\begin{equation}
[\calF^a(x), \, B'{}^b(y)]=-if^{abc}\calF^c(x)\delta(x^+-y^+).\label{F-B}
\end{equation}
With the help of \eqref{os2} and \eqref{os7}, \eqref{F-Phi} and 
\eqref{F-B} imply that $C^a\calF^a$ (anti)commutes with any fundamental 
field except for $\bC^b$:
\begin{equation}
\{ C^a(x)\calF^a(x), \, \bC^b(y) \} = \calF^b(x)\delta(x^+-y^+).
\end{equation}
We then find
\begin{equation}
\{ \QB-\hQB, \, \bC^b(y) \} = -\calF^b(y).              \label{Q-Q,bC}
\end{equation}
From \eqref{hQ-Phi} and \eqref{Q-Q,bC}, we have\footnote{
Note that we have not used $\calF^a=0$ in the above operator calculation.}
\begin{eqnarray}
\wightman{\bC^a(x_1)\hQB{}^2\bC^b(x_2)}
&=&\wightman{\{\bC^a(x_1),\,\hQB\}\{\hQB,\,\bC^b(x_2)\}}\nonumber\\
&=&\wightman{B^a(x_1)B^b(x_2)}=0,                       \label{hQ^2}\\
\wightman{\bC^a(x_1)\QB{}^2\bC^b(x_2)}
&=&\wightman{\{\bC^a(x_1),\,\QB\}\{\QB,\,\bC^b(x_2)\}}  \nonumber\\
&=&\wightman{(\partial_+\tB^a(x_1)-F^a(x_1))
             (\partial_+\tB^b(x_2)-F^b(x_2))}           \nonumber\\
&=&-\frac{D}{2(2\pi)^2}
    \frac{\delta^{ab}}{(x_1{}^+ - x_2{}^+ - i0)^2}      \label{Q^2}
\end{eqnarray}
owing to \eqref{fea2}.
\samepage{
\par
Thus we find that $\hQB$ is always nilpotent but the BRS Noether 
charge $\QB$ is {\it not\/} nilpotent for $D\not=0$.  The situation 
encountered here is quite similar to that of the bosonic string of 
Kato and Ogawa,\cite{KO} who analyzed the BRS-formulated CGQG and 
found that the BRS Noether charge is not nilpotent for $D\not=26$.  
We criticized their claim by showing that the exact solution to 
CGQG is BRS invariant and that there is another BRS generator which 
is always nilpotent.\cite{AN-KO} 
The result obtained in the present paper shows that similar phenomenon 
also occurs even {\it in a gauge theory}.  Thus we should conclude 
that the derivation of the critical dimension based on BRS anomaly 
is almost meaningless.
\par
Finally, we note that nothing anomalous happens for the FP-ghost 
number current in the present model. 
\vskip30pt
%
%
%
\section{Yang-Mills theory}
It is straightforward to generalize the results obtained in the 
previous sections to the case of the Yang-Mills theory. 
The Lagrangian density of Yang-Mills theory is given by \eqref{lag} 
plus the $\tB^2$ term as follows:
\begin{eqnarray}
\lag &=& 
 \tB^a(\partial_-A_+{}^a-\partial_+A_-{}^a-f^{abc}A_+{}^bA_-{}^c)
 - \frac{g^2}{2}\tB^a \tB^a \nonumber\\
 &&
 + B^a A_-^a + i\bC^a(\delta^{ab}\partial_- + f^{acb}A_-{}^c)C^b 
 + i\psi_M{}^\dagger(\partial_- - iA_-{}^aT^a)\psi_M,   \label{lagYM} \\
&&\nonumber
\end{eqnarray}
where $g$ is a coupling constant. 
Since the BF theory is trivially recovered as $g\to0$ in \eqref{lagYM}, 
both the conventional Yang-Mills theory and the BF theory are described 
by \eqref{lagYM}. Indeed, as long as $g\not=0$, we can make the field redefinition
\begin{equation}
\tB^a \to \tB^a 
+ \frac{1}{g^2}(\partial_-A_+{}^a-\partial_+A_-{}^a
                         -f^{abc}A_+{}^bA_-{}^c)      \label{redef}
\end{equation}
to obtain the conventional Lagrangian density for Yang-Mills theory
involving the $F_{-+}{}^2$ term with the extra $\tB^2$ term which 
is completely decoupled with the other fields. One should note that 
in the light-cone gauge \eqref{redef} is essentially a linear 
transformation, hence this field redefinition causes only a trivial
modification to the theory.
\par
Now, let us solve the model defined by \eqref{lagYM}. Field equations 
are the same as \eqref{fe1}--\eqref{fe3} except for \eqref{fe2} with 
$\varPhi=A_+{}^a$, which is replaced by
\begin{equation}
\partial_-A_+{}^a - g^2 \tB^a = 0.                    \label{fe-AtB}
\end{equation}
From \eqref{fe-AtB} and \eqref{fe2} with $\varPhi=\tB^a$, we obtain 
\begin{equation}
\partial_-{}^2 A_+=0.                                 \label{fe-A}
\end{equation}
Since the canonical (anti)commutation relations are the same as 
in the BF theory, two-dimensional (anti)commutation relations 
\eqref{os1}--\eqref{os3} are preserved in the present theory.
}
\vfill\eject\noindent%
\samepage{
To construct the two-dimensional commutation relation between 
$A_+{}^a(x)$ and $A_+{}^b(y)$, we set up the following Cauchy 
problem:
\begin{eqnarray}
\partial_-{}^y\,[ A_+{}^a(x), \, A_+{}^b(y) ] 
   &=& i g^2 \delta^{ab} \delta(x^+ - y^+),     \label{Cauchy-AA-1}\\
{}[ A_+{}^a(x), \, A_+{}^b(y) ]|_{y^0=x^0}
   &=& 0,                                                             \label{Cauchy-AA-2}
\end{eqnarray}
where \eqref{Cauchy-AA-1} is obtained from \eqref{fe-AtB} with 
\eqref{os1}. Then, the solution to \eqref{Cauchy-AA-1} with 
\eqref{Cauchy-AA-2} is uniquely given by
\begin{equation}
[ A_+{}^a(x), \, A_+{}^b(y) ] 
   = -i g^2 \delta^{ab}(x^- - y^-)\delta(x^+ - y^+).    \label{AA}       
\end{equation}
In the same way as LGBF, all other (anti)commutators vanish 
if the B-field $B^a$ is not involved. Furthermore, it is 
straightforward to show that \eqref{os5} for 
$\varPhi^a=\tB^a,\,C^a,\,\bC^a$ and \eqref{os6} remain unchanged. 
As for \eqref{os4}, it is modified as follows:
\begin{equation}
[B^a(x), \, A_+{}^b(y)]
    = i (\delta^{ab}\partial_+ + f^{acb}A_+{}^c(x))\delta(x^+-y^+)
     -i g^2 f^{acb}\tB^c(x) (x^--y^-)\delta(x^+-y^+).   \label{BA1}
\end{equation}
It is possible to rewrite \eqref{BA1} into
\begin{equation}
[B^a(x), \, A_+{}^b(y)]
    = i (\delta^{ab}\partial_+{} + f^{acb}A_+{}^c(y))\delta(x^+-y^+)
                                                        \label{BA2}
\end{equation}
which is trivially consistent with the field equation 
$\partial_-B^a=0$. Here, we have used the operator identity 
\begin{equation}
 \Big(A_+{}^a(x) - A_+{}^a(y) - g^2\tB^a(x)(x^- - y^-)\Big)
 \delta(x^+ - y^+) = 0,                         \label{AAtB-identity}
\end{equation} 
which is obtained as a trivial solution to the Cauchy problem
$\partial_-{}^xf(x,y)=0$ with $f(x,y)|_{x^0=y^0}=0$. 
Using \eqref{BA2}, we can show that \eqref{os5} for $\varPhi^a=B^a$ 
is satisfied. 
\par
It is, now, straightforward to obtain the operator subalgebra 
consisting of fields other than $\tB$. For that purpose, 
we have only to substitute $\frac{1}{g^2}\partial_- A_+$ to $\tB$ 
in \eqref{fe3}, \eqref{BA1} and \eqref{AAtB-identity}, 
and omit two-dimensional commutation relations involving $\tB$.  
Then, we find the resultant operator algebra is exactly the same 
as that in the $F_{-+}{}^2$ theory.
\par
Next, we consider Wightman functions. In order to make results 
useful also in the $F_{-+}{}^2$ theory, we avoid to use $B'{}^a$ 
defined by \eqref{B'}. 
In addition to \eqref{W-AtB}--\eqref{W-psiBpsi}, in which $B'{}^b$ 
is replaced by $B^b$, we obtain the following nonvanishing 2-point 
and 3-point functions 
\begin{eqnarray}
\hspace*{-30pt}&&\wightman{A_+{}^a(x_1)A_+{}^b(x_2)}
  =-\frac{g^2}{2\pi}\delta^{ab}
    \frac{x_1{}^--x_2{}^-}{x_1{}^+ - x_2{}^+ - i 0}, \\
\hspace*{-30pt}&&\wightman{A_+{}^a(x_1)B^b(x_2)}
  = -\frac{1}{2\pi}\delta^{ab}\frac{1}{(x_1{}^+ - x_2{}^+ - i0)^2}, \\ 
\hspace*{-30pt}&&\wightman{A_+^a(x_1)B^b(x_2)A_+^c(x_3)}
  = g^2f^{abc}(x_1{}^- - x_3{}^-) \varphi_3(x_1{}^+,x_2{}^+,x_3{}^+).  \label{ABA} \\
\hspace*{-30pt}&&\wightman{A_+^a(x_1)B^b(x_2)B^c(x_3)}
  = \frac{1}{(2\pi)^2}f^{abc}
        \frac{1}{(x_1{}^+\!-\!x_2{}^+\!-\!i0)(x_2{}^+\!-\!x_3{}^+\!-\!i0)
                 (x_1{}^+\!-\!x_3{}^+\!-\!i0)}. \qquad
\end{eqnarray}
}
\vfill\eject\noindent%
Likewise, additional nonvanishing truncated $n$-point functions 
$(n\geqq4)$ are given by
\begin{eqnarray}
\hspace*{-15pt}
&&\twightman{A_+{}^a(x_1)B^{b_2}(x_2)\cdots 
             B^{b_{n-1}}(x_{n-1})A_+{}^c(x_n)}   \nonumber\\
\hspace*{-15pt}
&&\quad
 =(-1)^{n-1}g^2\,(x_1{}^- - x_n{}^-)\sum_{P(j_2,\ldots,j_{n-1})}^{(n-2)!}
   f(ab_{j_2}d_{j_2})f(d_{j_2}b_{j_3}d_{j_3})\cdots
   f(d_{j_{n-2}}b_{j_{n-1}}c)                    \nonumber\\
\hspace*{-15pt}
&&\hspace*{200pt}
    \times\varphi(x_1{}^+,x_{j_2}{}^+,\ldots,x_{j_{n-1}}{}^+,x_n{}^+), \\
\hspace*{-15pt}
&&\twightman{A_+{}^a(x_1)B^{b_2}(x_2)\cdots B^{b_n}(x_n)} \nonumber\\
\hspace*{-15pt}
&&\quad
 = \frac{(-1)^n}{(2\pi)^{n-1}}\cdot\frac{1}{2}\sum_{P(j_2,\ldots,j_n)}^{(n-1)!}
    f(ab_{j_2}d_{j_2})f(d_{j_2}b_{j_3}d_{j_3})\cdots
    f(d_{j_{n-2}}b_{j_{n-1}}b_{j_n})                      \nonumber\\
\hspace*{-15pt}
&&\hspace*{30pt}\times
  \frac{1}{(x_1{}^+\!-\!x_{j_2}{}^+\!-\!i0) 
           (x_{j_2}{}^+\!-\!x_{j_3}{}^+\!\mp\! i0) \cdots
           (x_{j_{n-1}}{}^+\!-\!x_{j_n}{}^+\!\mp\! i0)  
           (x_{j_n}{}^+\!-\!x_{j_2}{}^+\!\mp\!i0)} 
    \qquad\quad  \label{AB...B}
\end{eqnarray}
with \eqref{xxmp0}.\footnote{In \eqref{AB...B}, the factor $\frac{1}{2}$ 
arises because $(j_2,\,j_3,\,j_4,\,\ldots,\,j_n)$ and 
$(j_2,\,j_n,\,\ldots,\,j_4,\,j_3)$ give the same contribution to the sum.}
Since the above Wightman functions affect neither 
$\wightman{B^a(x_1)B^b(x_2)}$ 
nor $\wightman{B^a(x_1)\calF^b(x_2)}$, where $\calF^a$ is defined 
by the lhs of \eqref{fe3} with substituting 
$\tB=\frac{1}{g^2}\partial_-A_+$, the field-equation anomaly in the 
BF theory remains unchanged in the Yang-Mills theory. 
So does the discussion on the BRS invariance made in Sec.\ 4.
\vskip30pt
%
%
%
\section{Discussion}
In the present paper, we have constructed the exact solutions to 
the BRS-formulated two-dimensional BF and Yang-Mills theories 
in the light-cone gauge. 
We have found that the solution to LGBF is quite similar to that 
of the BRS-formulated two-dimensional quantum gravity in the 
conformal gauge.  
In particular, it exhibits field-equation anomaly for the B-field 
equation. Previously, we constructed the exact solution to the 
BRS-formulated two-dimensional BF theory 
{\it in the Landau gauge}.\cite{AN-BF} 
Although in that work we did not discuss field-equation anomaly, 
we can check, by using the explicit expressions for the Wightman 
functions presented there, that field-equation anomaly is {\it not\/} 
encountered even if Dirac fields are taken into account. 
Furthermore, it was found that there is no field-equation anomaly 
in two-dimensional nonlinear abelian gauge models,\cite{Ikeda} though 
they are not BRS-formulated and without matter fields.  
Thus our result is the first finding of field-equation anomaly 
{\it in the gauge theory}.
\par
In the local-gauge commutation relations \eqref{os4}--\eqref{os6}, 
we have set the variables of fields appearing in the rhs to coincide 
with that of $B$ in the lhs, since any fields in LGBF is independent 
of $x^-$.  These forms are convenient in calculating truncated 
$n$-point functions.  To be precise, however, this choice of 
variables is different from the original expression for the local 
gauge-commutation relation defined by Kanno and Nakanishi.\cite{KN}  
In Yang-Mills theory, in which $A_+$ is dependent on $x^-$, the 
local-gauge commutation relation for $A_+$ is correctly described 
by \eqref{BA2} without non-gauge term, but not by \eqref{BA1}.
\par
Throughout this paper, we have restricted ourselves to considering 
only the left-handed chiral Dirac field $\psi_M$ $(M=1,...,D)$ 
as matter fields for simplicity of description. 
\vfill\eject\noindent
In LGBF, it is possible to add the right-handed chiral Dirac field 
$\chi_{M'}$ $(M'=1,...,D')$ coupled with $A_+$ without any fundamental 
difficulty. Even in this case, the B-field $B$ is independent on 
$x^-$, hence the local-gauge commutation relations are satisfied 
without non-gauge terms.  
Therefore, we have $\wightman{B^a(x_1)B^b(x_2)}=0$ and 
$\wightman{B^a(x_1)\calF^b(x_2)}\not=0$ unless $D=0$ in the same way 
as in Sec.\ 3.  Truncated $n$-point function consisting only of 
$\tB$'s no longer vanishes, but is proportional to $D'$.  
Since it satisfies 
$\partial_+{}^{x_2}\twightman{\tB^{a}(x_1)\tB^{b}(x_2)}=0$, 
we have $\wightman{\tB^a(x_1)\calF^b(x_2)}=0$, that is, no extra 
field-equation anomaly appears.
\par
We emphasize that the BRS structure of LGBF is almost completely 
parallel to that of CGQG. The latter is essentially nothing but 
the bosonic string.
Although Kato and Ogawa\cite{KO} claimed, by analyzing CGQG with 
string boundary conditions, that the critical dimension $D=26$ 
could be derived from the nilpotency condition of the BRS Noether 
charge, the solution to CGQG itself is BRS invariant and a BRS 
generator nilpotent for any value of $D$ can be constructed owing 
the the existence of field-equation anomaly. 
In the present paper, we have found also in LGBF that the solution 
is BRS invariant and a BRS generator nilpotent for any value of 
$D$ exists while the BRS Noether charge is not nilpotent for 
$D\not=0$.  We have to conclude, therefore, that the logical basis 
of deriving the critical dimension from the nilpotency condition 
for the BRS Noether charge is quite questionable.
\vskip30pt
%
%
%

\end{document}